\documentclass[aps,prd,amssymb,final,nobibnotes,nofootinbib,twocolumn,superscriptaddress]{revtex4-2}
\usepackage{graphicx}
\usepackage{hyperref}
\usepackage{amsmath}
\usepackage{xcolor}
\usepackage{numprint}
\usepackage{float}
\usepackage{ulem}
\usepackage{physics}
\usepackage{mathtools}
\usepackage{dsfont}
\usepackage{subfigure}

\begin{document}
\title{Nonminimally coupled scalar field in Schwarzschild--de Sitter spacetime: Geodesic synchrotron radiation}

\author{Jo\~ao P. B. Brito}
\email{joao.brito@icen.ufpa.br} 
\affiliation{Programa de P\'os-Gradua\c{c}\~{a}o em F\'{\i}sica, Universidade 
		Federal do Par\'a, 66075-110, Bel\'em, Par\'a, Brazil}

\author{Rafael P. Bernar}
\email{rbernar@ufpa.br} 
\affiliation{Programa de P\'os-Gradua\c{c}\~{a}o em F\'{\i}sica, Universidade 
		Federal do Par\'a, 66075-110, Bel\'em, Par\'a, Brazil}

\author{Lu\'is C. B. Crispino}
\email{crispino@ufpa.br}
\affiliation{Programa de P\'os-Gradua\c{c}\~{a}o em F\'{\i}sica, Universidade 
		Federal do Par\'a, 66075-110, Bel\'em, Par\'a, Brazil}

\date{\today}
\begin{abstract}
We analyze the scalar radiation emitted by a source interacting with a nonminimally coupled scalar field in four-dimensional Schwarzschild--de Sitter spacetime. We obtain the emission probability using quantum field theory in curved spacetimes at tree level. We find that the source emits synchrotron-type radiation for orbits near the photon sphere for all allowed values of the parameters. We also find that the emitted power strongly depends on the coupling to the curvature scalar. In particular,  the previously observed enhancement in the contribution of lower multipoles to the emitted power in this spacetime with minimal coupling is absent when conformal coupling is considered.
\end{abstract}

\maketitle

\section{Introduction}
Radio and gravitational wave astronomy~\cite{EHT_sombra,EHT_sombra_SgrA,ligo1_2016,ligo2_2016} have reached important milestones for gravitational physics with their groundbreaking results. These results have not only provided strong support for the experimental validity of general relativity (GR), but have also drawn increasing attention to black hole (BH) physics.

Black holes provide one of the best strong gravity settings to extract information about the physics of spacetime. One reason is that BHs can be exact solutions of the nonlinear GR field equations. These solutions are reasonably simple, as they can be described by a small set of parameters.
Another important reason is that the strong field regime provided by these compact objects plays a crucial role in the theory of GR and alternative theories of gravity~\cite{wald_1984,berti_2015}.
Despite GR being an exceptionally effective theory of gravity, it is a classical theory and breaks down near the Planck scale. This fate is usually signaled by the presence of singularities~\cite{Penrose1965}, which cause a loss of predictability in certain regions of the spacetime.
A quantum theory of gravity is expected to circumvent many of those problems in classical GR.
However, the formulation of this quantum theory remains one of the outstanding problems in fundamental physics~\cite{Kiefer2005,feynman_1995}.

In the absence of a quantum theory of gravity, we commonly resort to a semiclassical approach~\cite{birrell_1982,parker_2009} in which fundamental fields are quantized in fixed classical backgrounds.
While this semiclassical framework is regarded as an approximation, it is believed that it can provide insights into the quantum nature of gravity.
Quantum field theory (QFT) has yielded significant results, such as the creation of particles by dynamic spacetimes~\cite{parker_1969} and the prediction that BHs radiate (Hawking radiation), raising the possibility of their disappearance through thermal evaporation~\cite{hawking_1974,hawking_1975}.

Furthermore, there exists experimental evidence indicating that our Universe is currently experiencing an accelerated expansion~\cite{riess_1998,perlmutter_1999,Planck2020}, which suggests the presence of a nonzero cosmological constant~\cite{carroll_2001}.
In particular, the period of exponential inflation postulated to explain the very beginning of our Universe, can be approximately described by the de Sitter (dS) solution~\cite{guth_1981}.
It is also conceivable that our Universe is approaching a dS solution at late times~\cite{Perlmutter2000} and there is no reason to believe that stellar-mass BHs (or supermassive ones, for that matter) would have disappeared in the late dS-dominated era by, for instance, thermal evaporation~\cite{Page1976}. More importantly, during inflation, density perturbations might have given rise to primordial BHs~\cite{Carr1975}, and their role in the early structure formation dynamics is still poorly understood.
In this way, it is interesting to investigate BH physics in asymptotically dS solutions of GR. A particular solution of this type is the Schwarzschild--de Sitter (SdS) metric, which describes the spacetime of a static and uncharged BH inside an asymptotically dS universe. This family of solutions is also very important in the proposed dS/CFT correspondence~\cite{strominger_2001}.

Within the semiclassical approach, we investigate interaction processes in which, for instance, a classical source (e.g., a pointlike particle orbiting a BH) excites the quantum field. This phenomenon is interpreted as radiation emission, which depends on the motion of the source. In particular, when the source is a particle moving along ultrarelativistic circular geodesics, the resulting emitted radiation is of the synchrotron type (excitation of high-frequency harmonics beamed at narrow angles bisected by the orbital plane), featuring the so-called geodesic synchrotron radiation. These phenomena are astrophysically interesting, as BHs are expected to be surrounded by accretion disks. The pioneering works on geodesic synchrotron radiation~\cite{misner_1972,misner_et_al_1972}, arose as an attempt to interpret the observational data of Weber's gravitational wave experiments~\cite{weber_1969}, in which a reasonable explanation was the emission of gravitational synchrotron modes inferred by a detailed analysis of the scalar field.
Nonetheless, it became apparent shortly thereafter that, in this particular scenario, the field spin holds significant importance~\cite{Ruffini1972}. The lower multipole modes, especially the quadrupole one, are the predominant contributors to the emitted power, even for most unstable orbits~\cite{bernar_2017}.
Even though Weber's results were not indeed replicated, the analysis of geodesic synchrotron radiation scenarios is important because (i) the radiation emitted by matter spiraling into BHs plays a substantial role in high-energy astrophysics, and (ii) the understanding of the geodesic synchrotron radiation and the influence of the spin of the field on the emitted power is interesting in its own right~\cite{bernar_2018}.

We can analyze the geodesic synchrotron radiation using QFT in curved spacetimes, in particular in spherically symmetric spacetimes~\cite{bernar_2020}. The series of papers using this formalism started with Ref.~\cite{crispino_2000}, in which the scalar radiation emitted from a source rotating around a Schwarzschild BH in stable orbits was presented (see also Ref.~\cite{crispino_2008} for both stable and unstable orbits and Ref.~\cite{moreira_2021} for an analytical account). Other interesting scenarios have been investigated. The source coupled to a massive scalar field was reported in Ref.~\cite{castineiras_2007}, showing that the field mass decreases the emitted power by the source. 
The scalar radiation emitted from a rotating source around a Reissner--Nordstr\"{o}m BH was investigated in Ref.~\cite{crispino_2009},
around a Kerr BH in Ref.~\cite{macedo_2012}, 
around a Bardeen BH in Ref.~\cite{bernar_2019}, 
and around a Schwarzschild--(anti-)de Sitter BH in Refs.~\cite{brito_2020,brito_2021,brito_2021Add}. 
The electromagnetic and gravitational emitted radiation was investigated in Refs.~\cite{castineiras_2005,bernar_2017,bernar_2018}. The present study is a generalization of the one reported in Ref.~\cite{brito_2020}, in which the emitted power was computed considering the minimally coupled\footnote{The minimal coupling to gravity is implemented by making the substitution $\partial_{\mu} \to \nabla_{\mu}$ in the flat spacetime field equations.} scalar field in SdS spacetime.

Although numerous studies in the literature address the minimal coupling of the scalar field to gravity, massless (free) quantum fields in flat spacetime possess the property of conformal invariance, whereas their extension via minimal coupling to curved backgrounds generally does not~\cite{chernikov_1968,penrose_2011}.
Moreover, one can consider a generalization to curved backgrounds of the equation for a massive scalar field in flat spacetime, namely
\begin{equation}
  (\nabla^{\mu}\nabla_{\mu} -\mu^{2} - \xi R) \Phi = 0, \label{eq:massive-scalar-wave-equation}
\end{equation}
where $\xi$ is a general coupling constant to the curvature scalar $R$. In this case, if we require that the associated Green's function of Eq.~(\ref{eq:massive-scalar-wave-equation}) locally reduces to the one associated with the massive scalar wave equation in flat spacetime, one arrives at the so-called conformal coupling $\xi=1/6$ (in four dimensions). This requirement on wave propagation is related to the equivalence principle~\cite{sonego_1993,grib_1995}.
We note that, in the massless case ($\mu=0$), the conformal coupling value makes Eq.~(\ref{eq:massive-scalar-wave-equation}) conformally invariant, irrespective of the value of $R$.
Therefore, one can consider a general value of the coupling $\xi$, which forms a one-parameter family of possible couplings, paying close attention to the special case of conformal coupling.

We use QFT in curved spacetime at tree level to compute the scalar radiation emitted by a scalar source moving along circular geodesics around a SdS BH. We consider the general setting of a nonminimally coupled scalar field.
The differences in observable quantities between the cases of minimal and nonminimal coupling arise in curved backgrounds with $R \neq 0$, which is the case for the spacetime of a SdS BH.  We show that our results are substantially influenced by the value of $\xi$. The rest of this paper can be summarized as follows. In Sec.~\ref{sec_SdS_black_hole}, we review some important aspects of the SdS spacetime. In Sec.~\ref{sec_field_quantization}, we study the dynamics and quantization of the nonminimally coupled scalar field in the physical region of the SdS spacetime. In Sec.~\ref{sec_scalar_radiation}, we present the prescription of scalar radiation computation using numerically obtained radial mode solutions. In Sec.~\ref{sec_results}, we show our selected results. In Sec.~\ref{Sec_remarks}, we present some final remarks. We adopt geometrized units ($ c = G = \hbar = 1$) and the metric signature $(-, +, +, +).$

\section{Schwarzschild--de Sitter black holes}
\label{sec_SdS_black_hole}
Kottler's metric~\cite{kottler_1918} is a spherically symmetric vacuum solution of the GR field equations:
\begin{equation}
\label{GR_field_eq}
R_{\mu \nu} = \Lambda g_{\mu \nu},
\end{equation}
where $R_{\mu \nu}$ and $g_{\mu \nu}$ denote the Ricci and metric tensor components, respectively, and the Ricci scalar is defined by $g^{\mu \nu} R_{\mu \nu} \equiv R.$ From Eq.~\eqref{GR_field_eq}, we have $R=4\Lambda.$ The SdS spacetime is described by two (positive) parameters: the cosmological constant $\Lambda$ and the central BH geometric mass $M.$ The SdS line element is given by~\cite{stuchlik_1999}
\begin{equation}
\label{SdS_line_element}
ds^2 = -f_{\Lambda}(r)dt^2 + \frac{dr^2}{f_{\Lambda}(r)} + r^2(d\theta^2 + \sin^2 \theta d\phi^2),
\end{equation}
with
\begin{equation}
\label{f}
f_{\Lambda}(r) \equiv 1 - \frac{2 M}{r} - \frac{\Lambda}{3}r^2.
\end{equation}
The metric given by Eq.~\eqref{SdS_line_element} describes physics for radii $r$ between the BH horizon $r_h$ and the cosmological horizon $r_c$~\cite{akcay_2011,stuchlik_1999,rindler_2006}. The radial positions $r_h$ and $r_c$ are positive solutions of the third-degree polynomial $f_{\Lambda}(r) = 0$, and delimit a physical static region of the spacetime, which exists only if
\begin{equation}
\label{lambda_interval}
0\leq \Lambda < 1/9M^2.
\end{equation}
The lower limit of Eq.~\eqref{lambda_interval} corresponds to Schwarzschild spacetime, and the upper limit corresponds to the extreme SdS spacetime; in the latter case, the radial positions of the two horizons coincide with that of the photon sphere, $r_0 \equiv 3M.$ For $\Lambda > 1/9M^2,$ the physical region disappears and the spacetime presents no horizons. 
In Sec.~\ref{sec_field_quantization}, we review the quantization of the scalar field in the physical region of the SdS spacetime.

We shall consider a source rotating around a SdS BH. The existence region of timelike circular geodesics is given by (see Ref.~\cite{brito_2020})
\begin{equation}
\label{circular_range}
3M < r \leq \left( \frac{3M}{\Lambda} \right)^{1/3} \equiv r_{\mathrm{max}},
\end{equation}
and the angular velocity $\Omega$ of such orbits is expressed as
\begin{equation}
\label{angular_velocity}
\Omega \equiv \frac{d \phi}{dt} = \sqrt{\frac{M}{r^3} - \frac{\Lambda}{3}},
\end{equation}
which goes to zero at $r = r_{\mathrm{max}}.$
We note that the maximum of $f_{\Lambda}(r)$ is found at $r_{\mathrm{max}}$ [$f_{\Lambda}(r_{\mathrm{max}})  = 1- \sqrt[3]{9 M^2 \Lambda}$], where the BH attraction is balanced by the cosmological constant repulsion and the angular (transverse) tidal force changes sign~\cite{vandeev_2021}. Stable circular geodesics exist only for $\Lambda \leq (64/9)\times10^{-4} M^{-2}.$

\section{Nonminimally coupled scalar field}
\label{sec_field_quantization}
The following action can be used to derive the dynamics of a massless nonminimally coupled scalar field $\Phi(x):$
\begin{equation}
\label{action}
S =-\frac{1}{2}\int d^4x \sqrt{-g} \left[ \nabla_{\mu} \Phi(x) \nabla^{\mu} \Phi(x) + \xi R \Phi(x)^2 \right],
\end{equation}
where $g = - r^4 \sin^2 \theta$ is the metric determinant and $\xi$ is the dimensionless coupling constant where $\xi = 0$ corresponds to the so-called minimal coupling and $\xi = 1/6$ corresponds to the conformal coupling (in which the field theory is invariant under conformal transformations of the metric).

For the massless scalar field ($\mu=0$), Eq.~\eqref{eq:massive-scalar-wave-equation} can be written
as\footnote{Equation~\eqref{KG} has the same form as the equation of motion of the minimally coupled massive scalar field with mass parameter $\mu^2 \equiv \xi R$~\cite{higuchi_1987}.}
\begin{equation}
\label{KG}
\frac{1}{\sqrt{-g}}\partial_{\mu} \left[ \sqrt{-g} g^{\mu \nu} \partial_{\nu} \Phi(x) \right] - \xi R \Phi(x)= 0.
\end{equation}

Making use of the spacetime spherical symmetry, the solutions of Eq.~\eqref{KG} may be cast in the form
\begin{equation}
\label{Mode_positive}
u^k_{\omega \ell m}(x) = \sqrt{\frac{\omega}{\pi}} \frac{\Psi^k_{\omega \ell}(r)}{r} Y_{\ell m}(\theta, \phi) e^{- i \omega t} \hspace{0.4 cm} (\omega > 0).
\end{equation}
Equation~\eqref{Mode_positive} expresses the positive-frequency solution with respect to the timelike Killing vector field $\partial_t,$ where $Y_{\ell m}(\theta, \phi)$ represents the scalar spherical harmonics~\cite{NIST_handbook,gradshteyn_1980}.

The index $k$ in Eq.~\eqref{Mode_positive} indicates two types of modes. The $k = \mathrm{up}$ ($\mathrm{in}$) solutions denote modes purely incoming from the past event horizon $H^-_h$ (from the past cosmological horizon $H^-_c$), interacting with the effective potential, being partially reflected down to the future event horizon $H^+_h$ (to the future cosmological horizon $H^+_c$) and being partially transmitted to $H^+_c$ ($H^+_h$).

\begin{figure}
\includegraphics[scale=0.46]{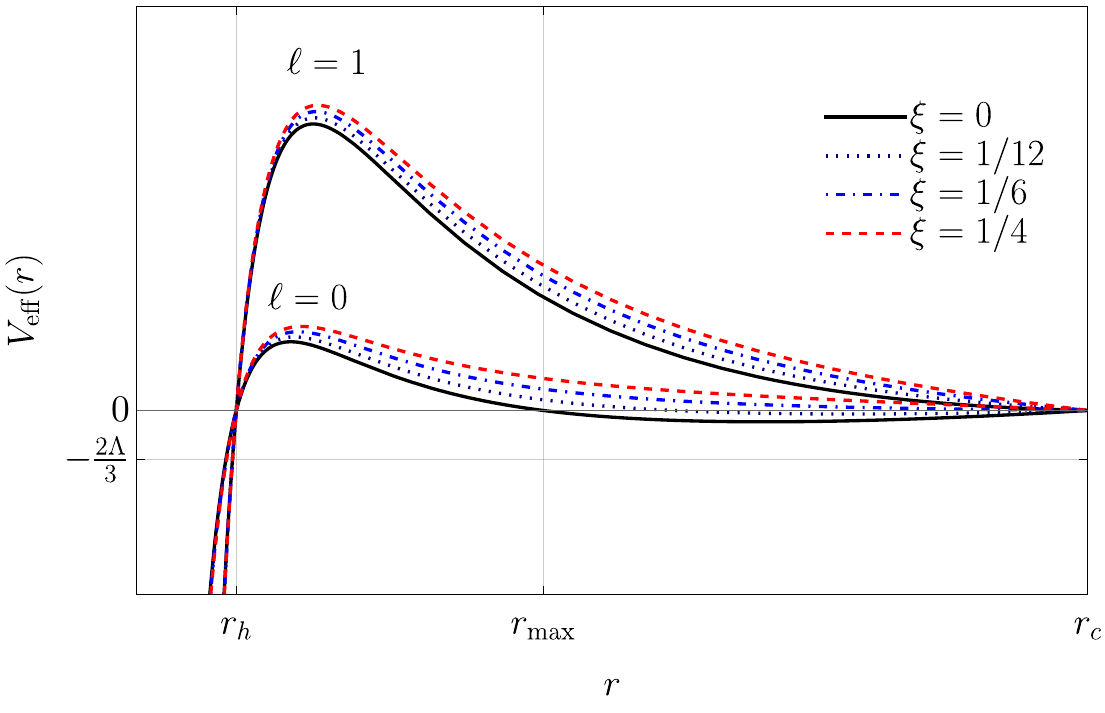}
\caption{The effective potential $V_{\mathrm{eff}},$ given by Eq.~(\ref{effective_potential}),  for two values of $\ell$ and different choices of the coupling constant $\xi$.}
\label{fig_effective_potential}
\end{figure}

We obtain the differential equation for the radial wave functions,
\begin{equation}
\label{radial}	
\left[ - f_{\Lambda}\frac{d}{dr} \left( f_{\Lambda}\frac{d}{dr}\right) + V_{\mathrm{eff}} \right] \Psi^k_{\omega \ell}(r) = \omega^2 \Psi^k_{\omega \ell}(r),
\end{equation}
by substituting Eq.~\eqref{Mode_positive} into Eq.~\eqref{KG}, with the effective potential defined as
\begin{equation}
\label{effective_potential}
V_{\mathrm{eff}}(r) \equiv f_{\Lambda}(r)\left( \frac{\ell(\ell+1)}{r^2} + \frac{2M}{r^3} - \frac{2 \Lambda}{3} +  \xi R\right),
\end{equation}
which vanishes at the position of both horizons, $r_h$ and $r_c.$ 
We note in passing that the effective potential diverges for $r \to \infty$, which is outside the physical region. This divergence does not occur if and only if the conformal coupling is considered ($\xi = 1/6$). 
In this case, $V_{\mathrm{eff}}(r \rightarrow \infty) \rightarrow -\ell(\ell+1)\Lambda/3$ (resembling the null geodesic potential with angular momentum $\ell+1/2$ for $\ell>>1$). From Fig.~\ref{fig_effective_potential}, we see that the potential barrier increases with $\xi.$ There is a point of maximum in the physical region for each $\ell$. For $\ell=0,$ the effective potential changes sign for some $\xi$ values (at $r = r_{\mathrm{max}},$ for $\xi = 0$), featuring a point of minimum inside the physical region. We note that, for a given $\Lambda,$ the potential is less sensitive to $\xi$ in the physical region near the event horizon and for large $\ell.$ For a given choice of $\xi,$ the potential barrier decreases as $\Lambda$ increases.

The approximate analytical solutions of Eq.~(\ref{radial}) on the vicinity of the horizons, for both $\mathrm{up}$ and $\mathrm{in}$ modes, are given by
\begin{equation}
\label{asymptotic_up}
\Psi^{\mathrm{up}}_{\omega \ell} = A_{\omega \ell}^{\mathrm{up}} \begin{cases}
 e^{i \omega r^*} + \mathcal{R}^{\mathrm{up}}_{\omega \ell}e^{-i\omega r^*}, & r \gtrsim r_h, \\
\mathcal{T}^{\mathrm{up}}_{\omega \ell}e^{i\omega r*}, & r \lesssim r_c,
\end{cases}
\end{equation}
and
\begin{equation}
\label{asymptotic_in}
\Psi^{\mathrm{in}}_{\omega \ell} = A_{\omega \ell}^{\mathrm{in}} \begin{cases}
 e^{-i \omega r^*} + \mathcal{R}^{\mathrm{in}}_{\omega \ell}e^{i\omega r^*}, & r \lesssim r_c, \\
\mathcal{T}^{\mathrm{in}}_{\omega \ell}e^{-i\omega r^*}, & r \gtrsim r_h,
\end{cases}
\end{equation}
where $r^*$ is the tortoise coordinate, implicitly defined by $dr^*=f_{\Lambda}(r)^{-1}dr$, and appropriate boundary conditions at $H_h$ and $H_c$ have been chosen. The quantity $A_{\omega \ell}^{k}$ denotes the overall normalization constants to be determined. The transmission (reflection) amplitudes are represented by the quantities $\mathcal{T}^{k}_{\omega \ell}$ ($\mathcal{R}^{k}_{\omega \ell}$). Note that we have considered a unitary outgoing flux at the past event horizon for $\mathrm{up}$ modes and a unitary incoming flux from the past cosmological horizon for the $\mathrm{in}$ modes. It is straightforward to show, from the Wronskian properties of the solutions, the conservation of the flux, expressed by 
\begin{equation}
\label{flux_conservation}
\abs{\mathcal{T}^k_{\omega \ell}}^2 + \abs{\mathcal{R}^k_{\omega \ell}}^2 = 1.
\end{equation}

The transmission coefficient in asymptotically flat spacetimes goes to zero in the limit $\omega \rightarrow 0$ and to unity for sufficiently high frequencies. This is only the case in SdS spacetime if $\xi \neq 0$ (nonminimal coupling)~\cite{crispino_2013}. 

The scalar field is canonically quantized in the usual way~\cite{birrell_1982,parker_2009,crispino_2000,higuchi_1987,ashtekar_1975}. We expand the quantum field operator $\hat{\Phi}(x)$ in terms of positive and negative frequencies, i.e.,
\begin{equation}
\label{field_expansion}
\hat{\Phi}(x) = \sum_{k,\ell,m} \int_0^{\infty} d\omega \left[u^{k}_{\omega \ell m}(x)\hat{a}^{k}_{\omega \ell m} + u^{k *}_{\omega \ell m}(x) \hat{a}^{k \dagger}_{\omega \ell m} \right],
\end{equation}
in which the coefficients $\hat{a}^{k}_{\omega \ell m}$ stand for annihilation operators and those $\hat{a}^{k \dagger}_{\omega \ell m}$ stand for creation operators.

We orthonormalize the modes $u^k_{\omega \ell m}(x)$ by considering the inner product~\cite{birrell_1982}
\begin{equation}
\left(\Phi, \Psi \right) \equiv i \int_{\Sigma} d\Sigma^{\mu}\left[ \Phi^* \left(\nabla_{\mu} \Psi \right) - \Psi \left( \nabla_{\mu} \Phi^* \right) \right],
\label{inner_product}
\end{equation}
where $d\Sigma^{\mu} = d\Sigma n^{\mu},$ with $n^{\mu}$ being a future-directed unit vector orthogonal to a Cauchy surface $\Sigma$ (e.g., the $t=\text{constant}$ hypersurface $\Sigma_t$). Since $\hat{\Phi}$ and $\hat{\Psi}$ satisfy Eq. (\ref{KG}), the inner product (\ref{inner_product}) is independent of the particular Cauchy hypersurface $\Sigma$ used~\cite{hawking_1973,parker_2009}. By requiring the orthogonality conditions
\begin{equation}
\label{u_orthogonality}
\left(u^{k}_{\omega \ell m},u^{k'}_{\omega' \ell' m'} \right) = \delta_{k k'} \delta_{\ell \ell'} \delta_{m m'} \delta(\omega - \omega')
\end{equation}
and
\begin{equation}
\label{u_ortogonality_null}
\left( u^k_{\omega \ell m},u^{k'*}_{\omega' \ell' m'} \right) = \left( u^{k*}_{\omega \ell m},u^{k'}_{\omega' \ell' m'} \right) = 0,
\end{equation}
one can readily show that the creation and annihilation operators satisfy the following nonvanishing commutation relations:
\begin{equation}
\label{comutation_relation}
\left[\hat{a}^k_{\omega \ell m},\hat{a}^{k' \dagger}_{\omega' \ell' m'} \right] =  \delta_{k k'} \delta_{\ell \ell'} \delta_{m m'} \delta(\omega - \omega').
\end{equation}

The vacuum state is defined as
\begin{equation}
\label{vacuum}
\hat{a}^{k}_{\omega \ell m} \ket{0} \equiv 0, \hspace{1 cm} \forall \hspace{0.3 cm} (k, \omega, \ell, m),
\end{equation}
and the one-particle-state, described by the quantum numbers $\ell,$ $m,$ and energy $\omega,$ is written as
\begin{equation}
\label{estad_uma_partic}
\hat{a}^{k \dagger}_{\omega \ell m} \ket{0} = \ket{k; \omega \ell m}.
\end{equation}

From Eqs.~(\ref{inner_product})--(\ref{u_ortogonality_null}) and the radial modes differential equation \eqref{radial}, written in terms of the tortoise coordinate, we obtain the modulus of the overall normalization constants, namely 
\begin{equation}
\label{normalization_constant}
\abs{A^{\mathrm{up}}_{\omega \ell m}} = \abs{A^{\mathrm{in}}_{\omega \ell m}} = \frac{1}{2 \omega}.
\end{equation}

\section{Scalar radiation and emitted power}
\label{sec_scalar_radiation}
We consider a nonminimally coupled scalar field interacting with a classical source. The radiation process can be analyzed by calculating the transition amplitude from the vacuum-state to the one-particle-state, defined in Eqs.~\eqref{vacuum} and \eqref{estad_uma_partic}, respectively.

The source is a pointlike particle moving along a timelike circular geodesic at the radial position $r=R_0$ with angular velocity $\Omega$ in the equatorial plane ($\theta = \pi/2$ and $\dot{\theta}=0$). Thus, it is described by the following normalized current:
\begin{equation}
\label{current}
j(x) = \frac{\sigma}{\sqrt{-g} v^0} \delta(r-R_0) \delta(\theta - \pi/2)\delta(\phi - \Omega t),
\end{equation}
where $\sigma$ is a constant. The quantity $v^0$ is the $t$-component of the particle's $4$-velocity, $v^{\mu}$, which is given by
\begin{equation}
\label{four_velocity}
v^{\mu} = \left(1,0,0,\Omega \right)/\sqrt{f_{\Lambda}(R_0) - R_0{}^2 \Omega^2},
\end{equation}
with $\Omega$ defined in Eq.~\eqref{angular_velocity}.
It is worth noting that $\int d\zeta^{(3)}j(x) = \sigma,$ where the hypersurface $\zeta^{(3)}$ is a Cauchy hypersurface.

The interaction action operator is denoted as
\begin{equation}
\label{interaction_action}
\hat{S}_{I} = \int d^4x \sqrt{-g} j(x) \hat{\Phi}(x),
\end{equation} 
where the magnitude of the interaction is determined by the constant $\sigma.$

At the lowest order in perturbation theory, the transition from the vacuum-state, as defined in Eq.~(\ref{vacuum}), to the one-particle-state, as defined in Eq.~\eqref{estad_uma_partic}, has the following amplitude~\cite{itzykson_1980}:
\begin{equation}
\label{probability_amplitude}
\mathcal{A}^{k; \omega \ell m}_{\mathrm{em}} = \bra{k; \omega \ell m} i \hat{S}_I \ket{0} = i \int d^4x \sqrt{-g} j(x) u^{k *}_{\omega \ell m}.
\end{equation}
The expansion of Eq.~\eqref{probability_amplitude} yields $\mathcal{A}^{k; \omega \ell m}_{\mathrm{em}} \propto \delta(\omega - m\Omega),$ indicating that only scalar quanta with $\omega_m \equiv m\Omega$ are emitted. Recall that $m$ is an integer, so that, for a given circular orbit, the emitted spectrum is discrete.

For fixed values of $k,$ $\ell,$ and $m,$ the (partial) emitted power is given by
\begin{equation}
\label{partial_power_implicit}
W^{k; \ell m}_{\mathrm{em}} = \int_{0}^{\infty} d\omega \omega \frac{\abs{\mathcal{A}^{k; \omega \ell m}_{\mathrm{em}}}^2}{T}.
\end{equation} 
We consider the case in which the source radiates during the whole range of coordinate time $t,$ with $-\infty < t < \infty.$ Thus, we can write $T=\int dt = 2 \pi \delta(0)$~\cite{breuer_1975,crispino_1998}. As a result, Eq.~\eqref{partial_power_implicit} becomes
\begin{equation}
\label{partial_power}
W^{k;\ell m}_{\mathrm{em}} = 2 \sigma^2 \omega_m^2 \left(1-\frac{3M}{R_0} \right) \abs{\frac{\Psi^k_{\omega_m \ell}}{R_0}}^2 \abs{Y_{\ell m}\left(\frac{\pi}{2},0 \right)}^2,
\end{equation}
where the dependence in $R_0$ can be written in terms of $\Omega$ by using Eq.~\eqref{angular_velocity}. 
We can see from Eq.~\eqref{partial_power} that $W^{k;\ell m}_{\mathrm{em}}$ vanishes for $R_0 \to 3M$. At $R_0 \to r_{\mathrm{max}}$, the partial emitted power also vanishes due to $\Omega$ being zero, which means $\omega_{m}=0$.

We note that one can regard the $\xi R$ term in Eqs. \eqref{action} and \eqref{KG} as a mass term $\mu^2$. In Schwarzschild spacetime, the partial emitted power exhibits a step-function behavior $\Theta(\omega-\mu)$~\cite{castineiras_2007}. This is because the emitted quanta associated with the in-modes must have frequency restricted by $\omega \geq \mu$, i.e., the quantum energy cannot be less than the mass parameter. The up-modes have no such restriction~\cite{castineiras_2002}. However, in SdS, the in-modes are the ones coming from the past cosmological horizon, which implies no restriction on their frequencies. This is because the effective potential, given by Eq.~\eqref{effective_potential}, vanishes at the cosmological horizon. In contrast, the effective potential in the Schwarzschild case tends to a constant value at infinity.

Summing the contributions of all partial powers for both $\mathrm{in}$ and $\mathrm{up}$ modes, we obtain the total emitted power, i.e.,
\begin{equation}
\label{total_power}
W_{\mathrm{em}} = \sum_{\ell=1}^{\infty} \sum_{m=1}^{\ell} W^{\ell m}_{\mathrm{em}},
\end{equation}
where
\begin{equation}
\label{sum_in_up_Tot_power}
W^{\ell m}_{\mathrm{em}} =  W^{\mathrm{in};\ell m}_{\mathrm{em}} + W^{\mathrm{up};\ell m}_{\mathrm{em}}.
\end{equation}

By solving Eq.~\eqref{radial} numerically in the physical region, we obtain the modes $\Psi^k_{\omega \ell}$ for any arbitrary position $r$ and frequency $\omega.$ Some key results are presented in the next section.

\section{Results}
\label{sec_results}
In this section, we analyze the emitted power for representative values of the cosmological constant $\Lambda$ and the coupling constant $\xi.$ In particular, we compare the results with minimal and conformal coupling. 
The plots range from $\Omega(r_{\mathrm{max}})\equiv \Omega_{\mathrm{min}}=0$ up to $\Omega(r_0)\equiv \Omega_{\mathrm{max}}.$ The quantity $\Omega_{\mathrm{max}},$ i.e., the angular velocity of the source at the photon sphere, is indicated by the vertical (gray) line in the figures.

We recall that the major contribution to the emitted power for orbits close enough to the photon sphere comes from the modes with $\ell=m.$ Note that the factor $\abs{Y_{\ell m}(\pi/2,0)}^2,$ on the right-hand side of Eq. \eqref{partial_power}, is maximum for $\ell=m,$ and decays exponentially as $m$ decreases from $m=\ell$ for a given large $\ell$. Considering $\ell=20$ and $M^2 \Lambda = 150^{-1},$ with $R_0 = 0.9 r_{\mathrm{max}},$ the contribution of the mode with $\ell=m$ to the emitted power is approximately $97 \%$ and it is larger than $97 \%$ for smaller $R_0.$ This dominance of the $\ell=m$ modes holds true for all allowed values of the parameters.

\begin{figure}[h!]
\center
\includegraphics[scale=0.45]{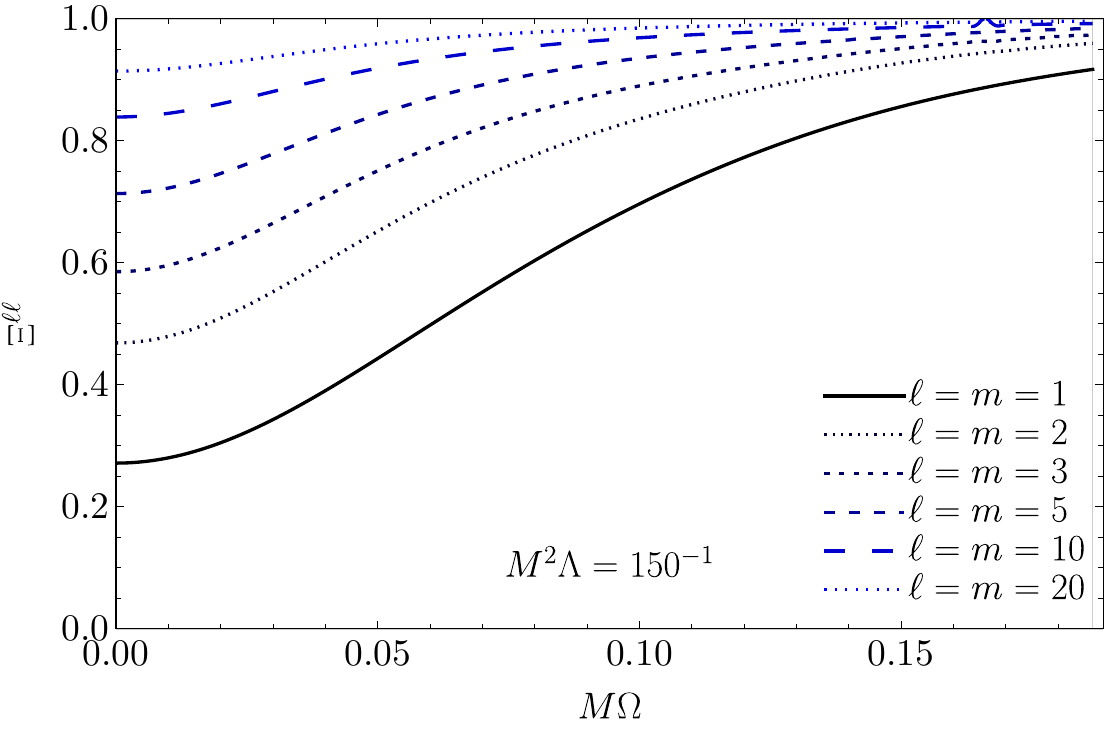}
\includegraphics[scale=0.45]{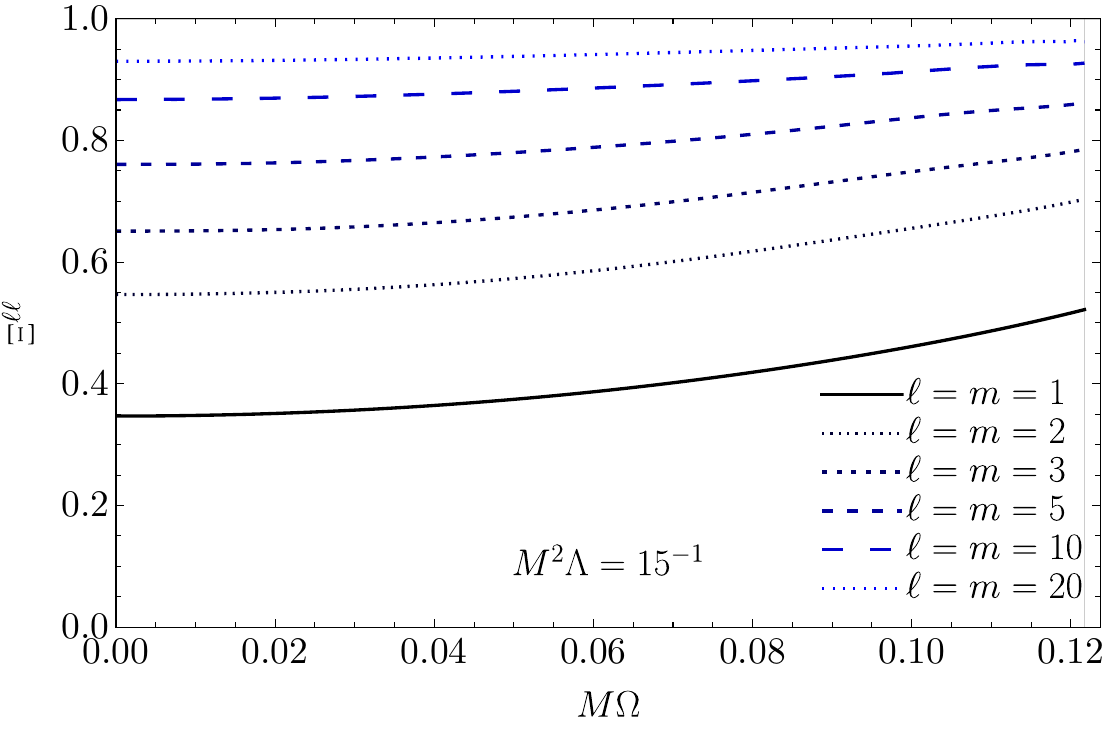}
\caption{The function $\Xi^{\ell m},$ given by Eq.~\eqref{Xi_ratio}, for $M^2 \Lambda = 150^{-1}$ (top) and $M^2 \Lambda = 15^{-1}$ (bottom), with some choices of $\ell=m.$}
\label{Ratio_xi_Tot_Lamb1}
\end{figure}

We define a quantity $\Xi^{\ell m}$ as the ratio between the emitted power with $\xi = 1/6$ and the one with $\xi = 0,$ namely
\begin{equation}
\label{Xi_ratio}
\Xi^{\ell m} \equiv \frac{W^{\ell m}_{\mathrm{em}}\Big\vert_{\xi = 1/6}}{W^{\ell m}_{\mathrm{em}}\Big\vert_{\xi = 0}}.
\end{equation}
The quantity $\Xi^{\ell m},$ given by Eq.~\eqref{Xi_ratio}, is illustrated in Fig.~\ref{Ratio_xi_Tot_Lamb1} for two representative values of $M^2 \Lambda$ and different choices of $\ell=m.$ For $M^2 \Lambda = 150^{-1},$ we see that $\Xi^{\ell m}$ decreases as $M \Omega$ decrease. The differences in emitted power between the minimal coupling case and the conformal one increase as we consider lower multipoles.
For $M^2 \Lambda = 15^{-1}$, these differences are much more pronounced across the entire $M\Omega$ range as one can see from the ratio $\Xi^{\ell m}$, especially for low values of $\ell=m$.
This is consistent with the behavior of the effective potential \eqref{effective_potential}, which increases with increasing $\xi,$ especially  for small $\ell=m$ and large $R_0.$ Figure~\ref{Ratio_xi_Tot_Lamb1} implies that the emitted power with $\xi=1/6$ can differ significantly from that with $\xi=0.$ The influence of $\xi$ is enhanced by increasing $\Lambda.$

\begin{figure}
\center
\includegraphics[scale=0.45]{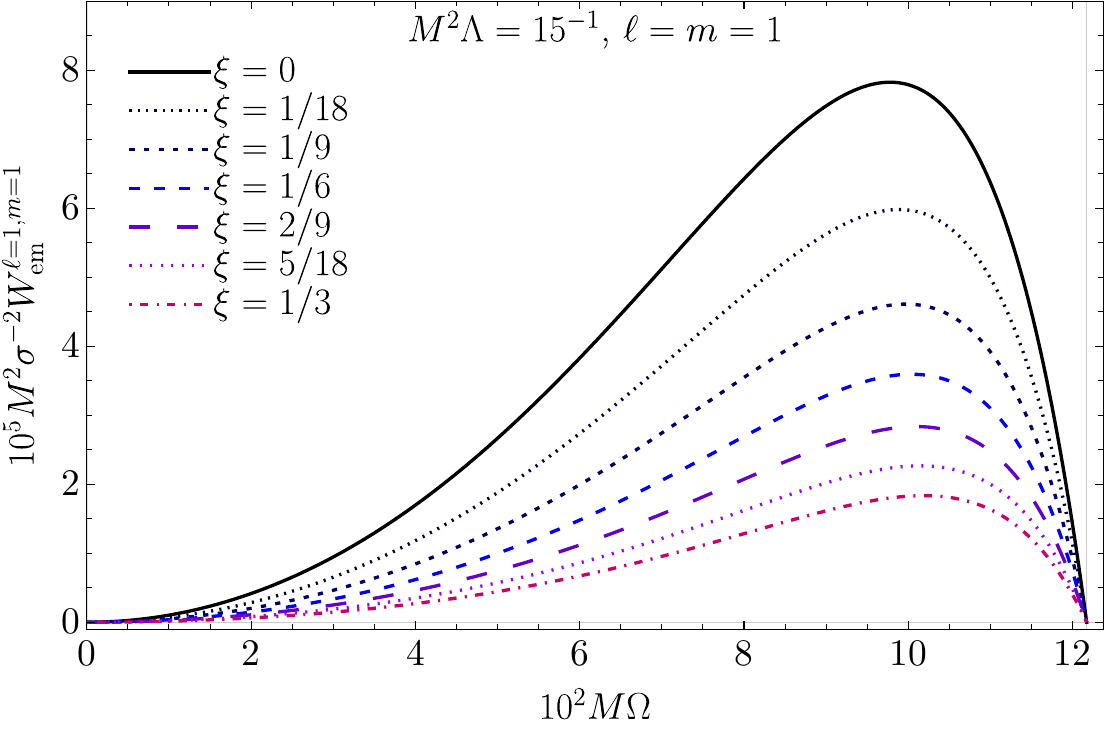}
\includegraphics[scale=0.45]{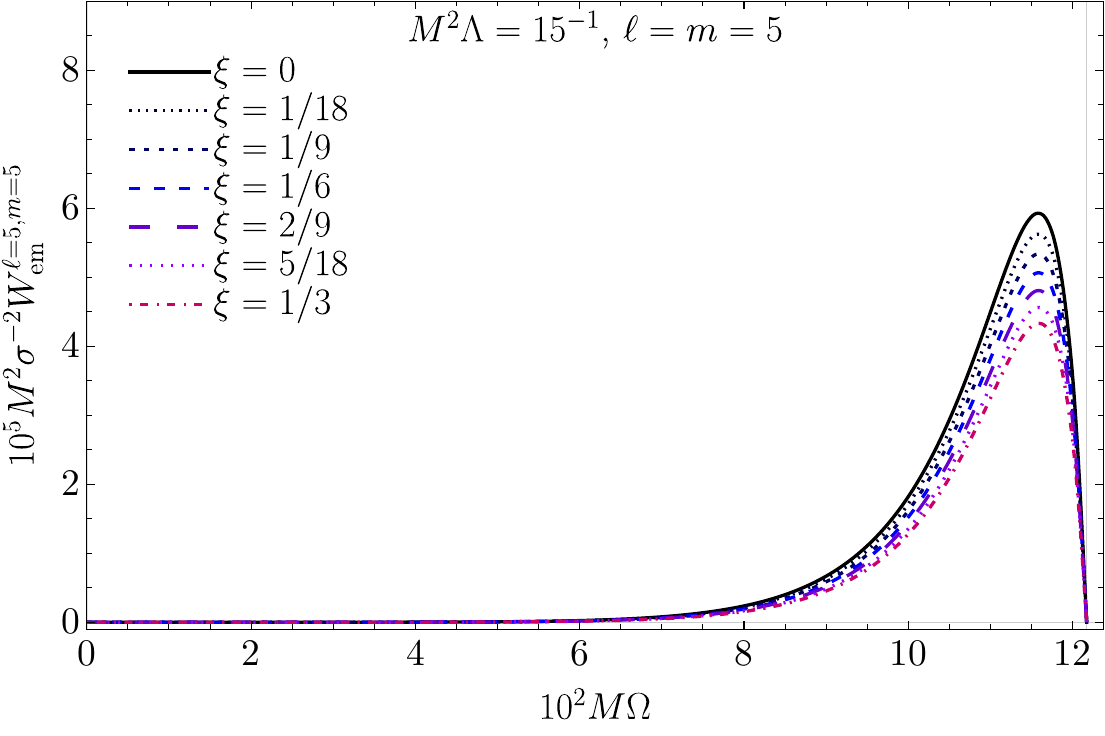}
\caption{The emitted power $W_{\mathrm{em}}^{\ell=1,m=1}$ (top) and $W_{\mathrm{em}}^{\ell=5,m=5}$ (bottom), given by Eq.~\eqref{sum_in_up_Tot_power}, as a function of $M\Omega$  for $M^2\Lambda = 15^{-1}$ and different choices of the coupling constant $\xi$.}
\label{Pot_Tot_Lamb1_15_l_1_m_1_x}
\end{figure}

Figure~\ref{Pot_Tot_Lamb1_15_l_1_m_1_x} shows the partial emitted power for $\ell=m=1$, $\ell=m= 5$ and different choices of the coupling constant $\xi.$ We see that the power is enhanced as the coupling constant decreases from $1/3$ to zero. 
We note that for higher multipoles, this enhancement becomes fainter, so that for $\ell=m=20$ the peak of the power with $\xi = 1/3$ is approximately $8\%$ smaller than the one with $\xi = 0.$

\begin{figure}[h!]
\center
\includegraphics[scale=0.45]{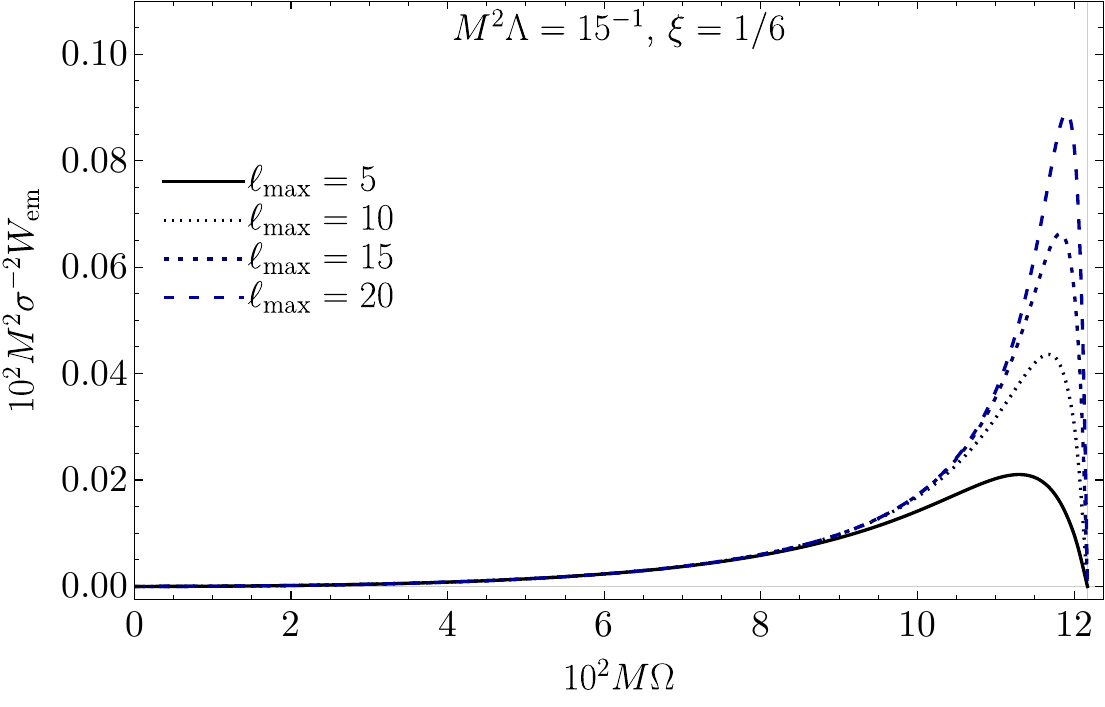}
\caption{The total emitted power $W_{\mathrm{em}}$ as a function of $\Omega,$ given by Eq.~\eqref{total_power}, for $M^2\Lambda = 15^{-1}$ and some choices of $\ell_{\mathrm{max}}.$}
\label{Pot_Tot_Lamb1_lmax}
\end{figure}

The total emitted power, given by Eq.~\eqref{total_power}, is plotted in Fig.~\ref{Pot_Tot_Lamb1_lmax} for a representative value of $M^2 \Lambda.$ The sum in $\ell$ was truncated to a chosen maximum value, $\ell_{\mathrm{max}}.$ We see that, in the vicinity of the photon sphere, the contribution of higher multipoles becomes predominant, indicating the emission of synchrotron-type radiation. Furthermore, as the coupling constant increases from zero, the total emitted power for each $\ell_{\mathrm{max}}$ decreases (see Ref.~\cite{castineiras_2007}).

\begin{figure*}
\center
\includegraphics[scale=0.40]{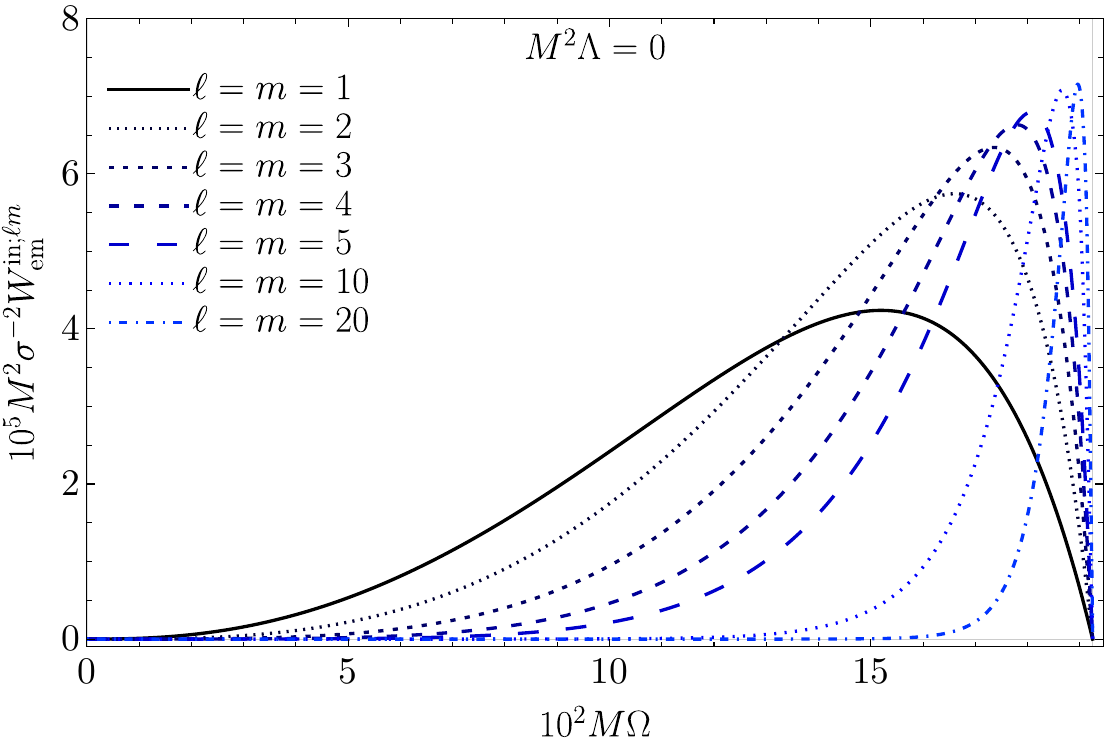} \hspace{0.2 cm}\vspace{0.3 cm}
\includegraphics[scale=0.40]{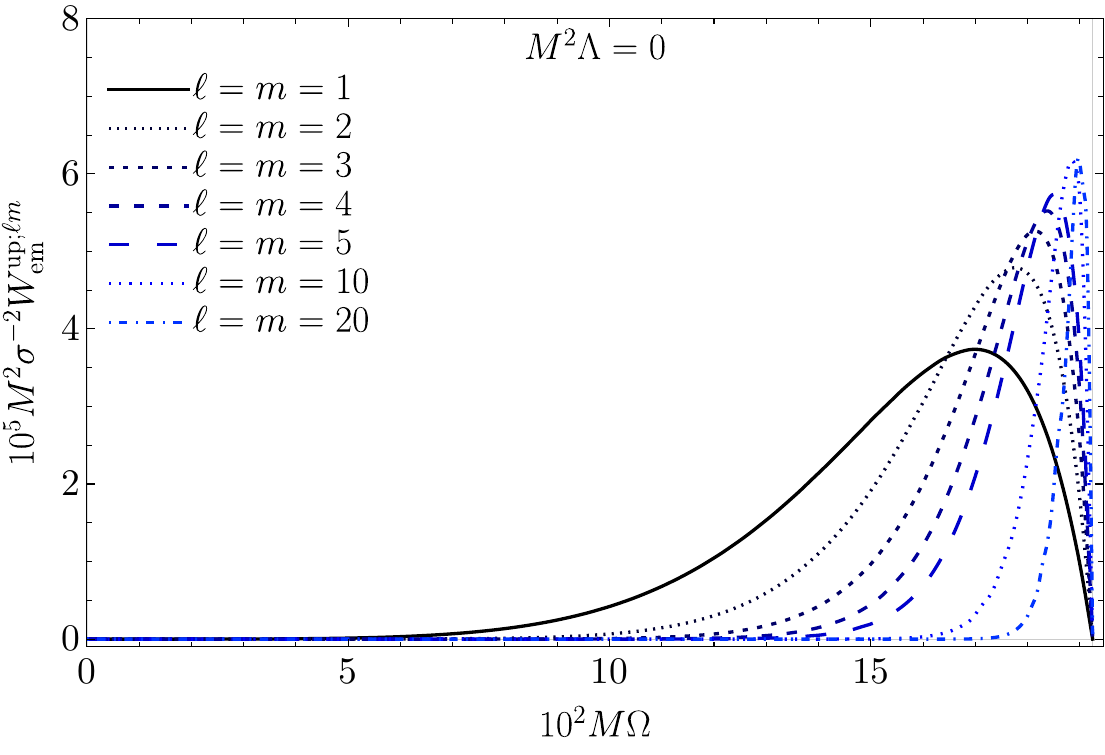} \vspace{0.2 cm}
\includegraphics[scale=0.40]{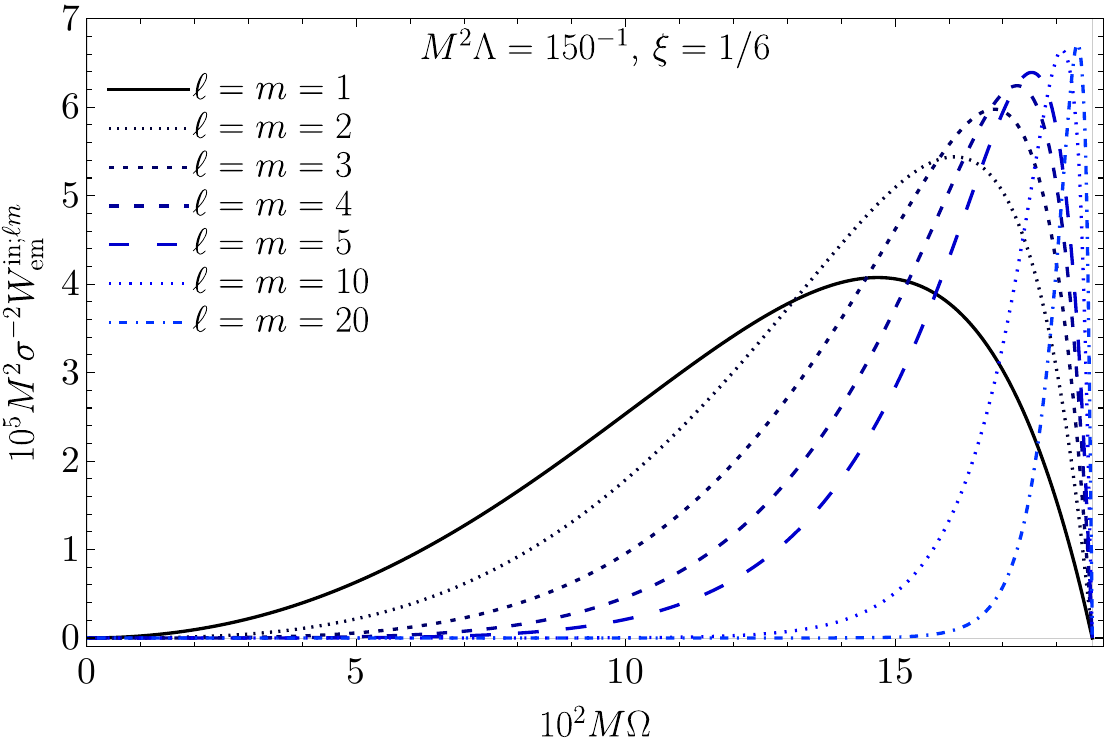} \hspace{0.2 cm}\vspace{0.3 cm}
\includegraphics[scale=0.40]{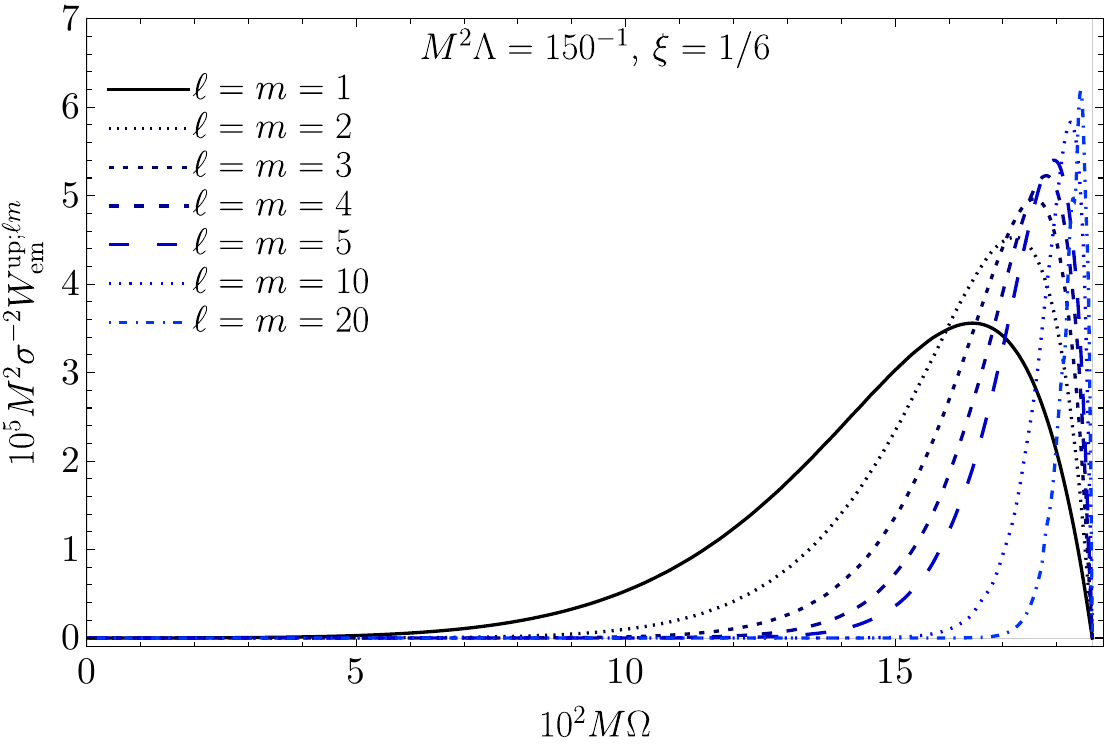} \vspace{0.2 cm}
\includegraphics[scale=0.40]{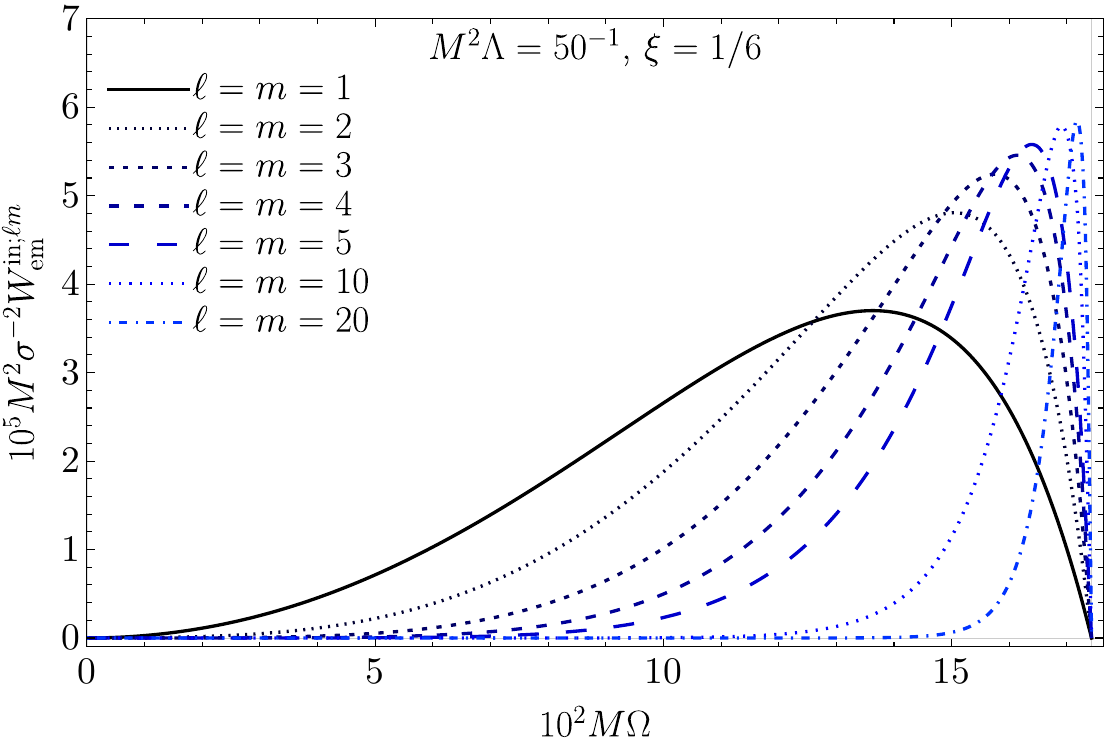} \hspace{0.2 cm}
\includegraphics[scale=0.40]{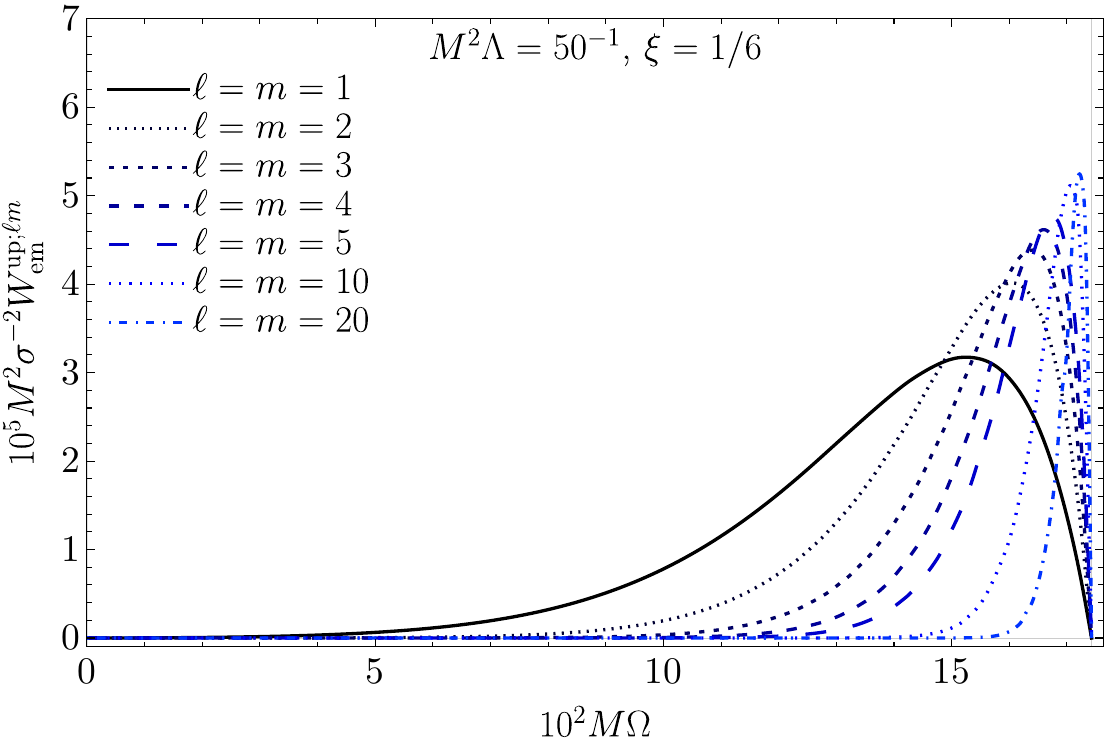} 
\includegraphics[scale=0.40]{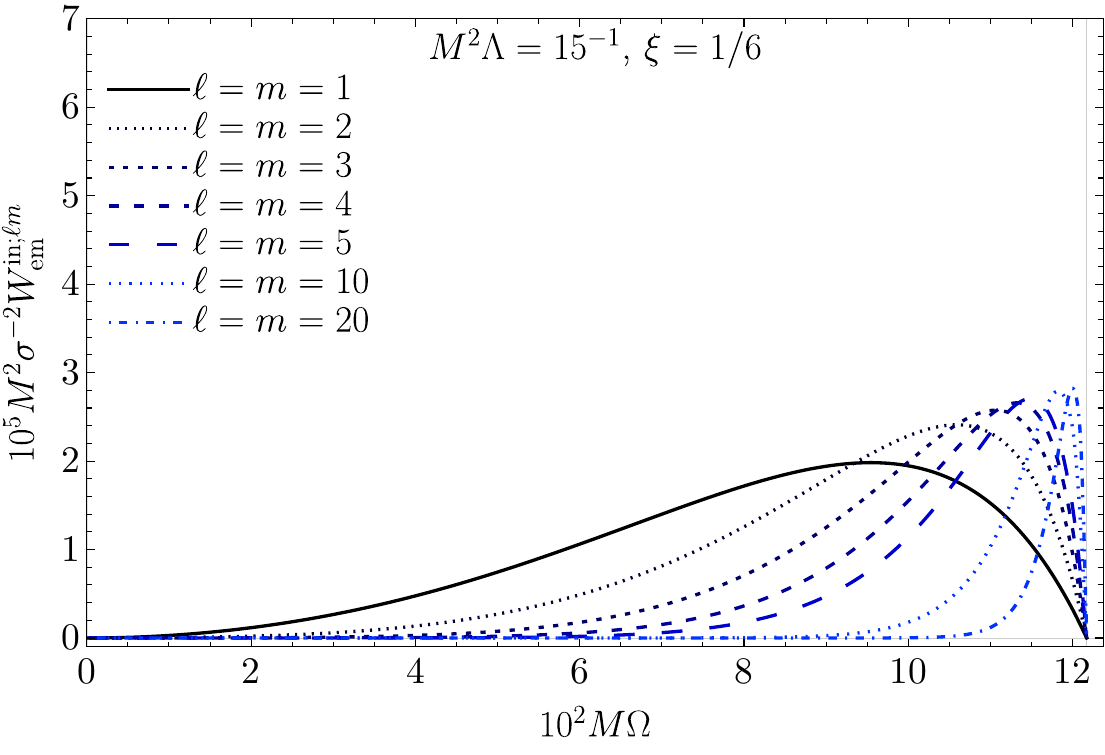} \hspace{0.2 cm}
\includegraphics[scale=0.40]{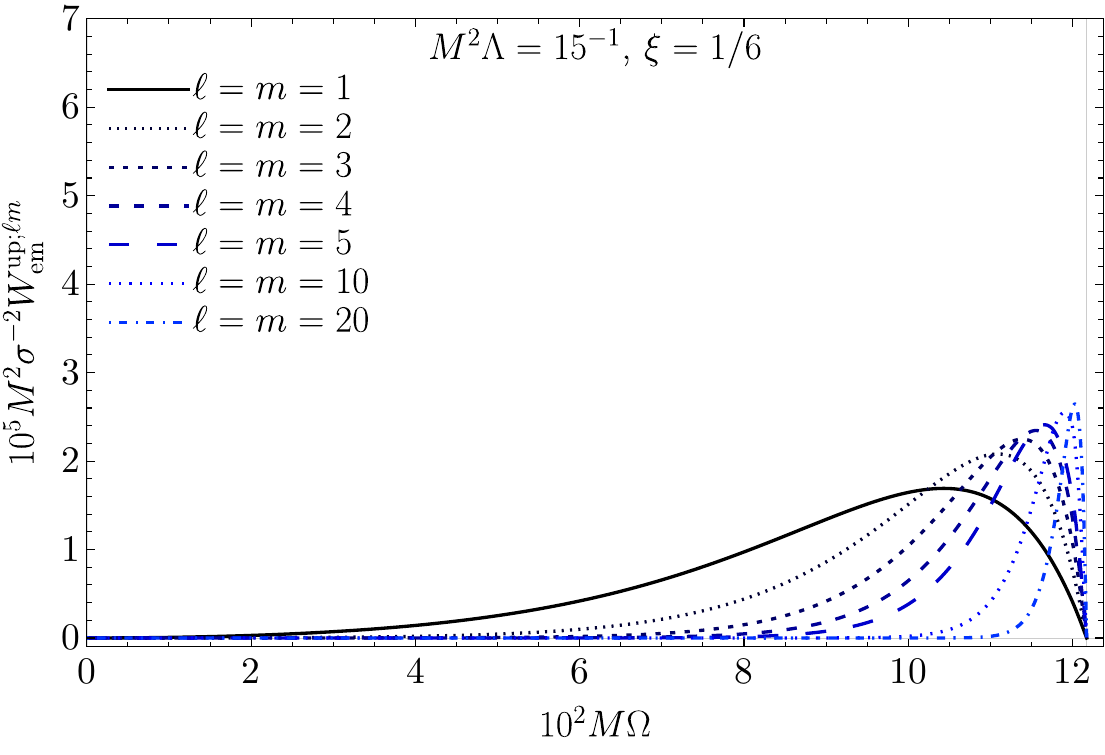} 
\caption{The emitted power $W^{k;\ell m}_{\mathrm{em}}$ with $\xi = 1/6$ as a function of $M\Omega,$ given by Eq.~(\ref{partial_power}), for modes $\mathrm{in}$ (left) and $\mathrm{up}$ (right), with some choices of $\ell=m.$ The plots are shown for four values of $M^2 \Lambda,$ from $M^2 \Lambda = 0$ to $M^2 \Lambda = 15^{-1}$.}
\label{Pot_Parc_IN_e_UP}
\end{figure*}

Figure~\ref{Pot_Parc_IN_e_UP} shows separately the contribution of the $\mathrm{in}$ and $\mathrm{up}$ modes to the emitted power with $\xi=1/6$, for some choices of $\ell=m$ and four values of $M^2 \Lambda,$ including the Schwarzschild case ($M^2 \Lambda = 0$).
The radial position of the peak of $W^{k;\ell m}_{\mathrm{em}}$ approaches $\Omega(r_0)$ as $\ell$ increases, exhibiting a characteristic behavior~\cite{crispino_2008,macedo_2012,bernar_2019,brito_2020}. 
Note that these results are quite different from the ones with minimal coupling (see Fig. 7 of Ref.~\cite{brito_2020}). In particular, the enhancement observed in the contribution of lower multipoles to the emitted power with minimal coupling is absent in the conformal coupling case.
\section{Final Remarks}
\label{Sec_remarks}
We have analyzed the scalar geodesic synchrotron radiation in  Schwarzschild--de Sitter spacetime using quantum field theory in curved spacetimes at tree level. In particular, using numerical computations, we have compared the results for the scalar emitted power minimally and nonminimally coupled to the curvature scalar in the entire allowed range of the parameter $\Lambda.$

Our results indicate that the emitted power depends on the coupling constant $\xi$. As a general rule of thumb, the role of the coupling here is to reduce the amount of emitted radiation, when compared to the minimal coupling case. These results complement other studies of the scalar field and are important for understanding dynamical processes in asymptotically de Sitter spacetimes (see, e.g., Refs.~\cite{kanti_2005,crispino_2013} and references therein).

We have also shown that the source emits synchrotron-type radiation while orbiting the black hole near the photon sphere for all allowed values of $\Lambda$ and $\xi.$ Considering a conformally coupled scalar field, we have also presented the total emitted power and the contribution of each $\mathrm{in}$ and $\mathrm{up}$ modes separately. In the conformal coupling case, the emitted power behavior, for a nonvanishing cosmological constant, is quite similar to the minimal coupling case with a vanishing cosmological constant.

\section*{Acknowledgments}
We are grateful to Funda\c{c}\~ao Amaz\^onia de Amparo a Estudos e Pesquisas (FAPESPA), Conselho Nacional de Desenvolvimento Cient\'ifico e Tecnol\'ogico (CNPq) and Coordena\c{c}\~ao de Aperfei\c{c}oamento de Pessoal de N\'ivel Superior (CAPES)--Finance Code 001, from Brazil, for partial financial support. This work has further been supported by the European Union's Horizon 2020 research and innovation (RISE) programme H2020-MSCA-RISE-2017 Grant No. FunFiCO-777740 and by the European Horizon Europe staff exchange (SE) programme HORIZON-MSCA-2021-SE-01 Grant No. NewFunFiCO-101086251.


\end{document}